# Chiral symmetry breaking for deterministic switching of perpendicular magnetization by spin-orbit torque

Hao Wu[1,*], John Nance[2], Seyed Armin Razavi[1], David Lujan[3], Bingqian Dai[1], Yuxiang Liu[1], Haoran He[1], Baoshan Cui[1], Di Wu[1], Kin Wong[1], Kemal Sobotkiewich[3], Xiaoqin Li[3], Gregory P. Carman[2], and Kang L. Wang[1,†]

[1]*Department of Electrical and Computer Engineering, and Department of Physics and Astronomy, University of California, Los Angeles, California 90095, USA*

[2]*Department of Mechanical and Aerospace Engineering, University of California, Los Angeles, California 90095, USA*

[3]*Department of Physics, and Center for Complex Quantum Systems, The University of Texas at Austin, Texas 78712, USA*

*Corresponding author. wuhaophysics@ucla.edu

†Corresponding author. wang@ee.ucla.edu

**Abstract:**

Symmetry breaking is a characteristic to determine which branch of a bifurcation system follows upon crossing a critical point. Specifically, in spin-orbit torque (SOT) devices, a fundamental question arises: how to break the symmetry of the perpendicular magnetic moment by the in-plane spin polarization? Here, we show that the chiral symmetry breaking by the antisymmetric Dzyaloshinskii–Moriya interaction (DMI) can induce the deterministic SOT switching of the perpendicular magnetization. By introducing a gradient of saturation magnetization or magnetic anisotropy, non-collinear spin textures are formed by the gradient of effective SOT strength, and thus the chiral symmetry of the SOT-induced spin textures is broken by the DMI, resulting in the deterministic magnetization switching. We introduce a strategy to induce an out-of-plane ($z$) gradient of magnetic properties, as a practical solution for the wafer-scale manufacture of SOT devices.

Symmetry, is a fundamental characteristic of a system that is preserved under some transformations. Symmetry breaking often results in important phenomena in physics, such as quantized conductance and superconductivity [1]. In spintronics, the symmetry of the perpendicular magnetization is preserved under the current induced spin-orbit torque (SOT) [2-4] and hence the SOT switching is non-deterministic. SOT originates from a vertical spin current ($J_s$) injection from an adjacent layer or interface with strong spin-orbit coupling by an in-plane (electric) charge current ($J_e$), via the way of spin Hall effect [5] or Rashba effect [6], and the spin polarization $\sigma$ is along the in-plane direction according to $J_s \propto \sigma \times J_e$; therefore, for the magnetization with perpendicular magnetic anisotropy (PMA) [7, 8], the spins with in-plane polarization [damping-like torque $\tau_{SOT} \propto m \times (m \times \sigma)$] cannot break the symmetry between perpendicular magnetizations of $\pm M_z$, where $m$ represents the unit vector of the magnetic moment, leading to the non-deterministic nature of SOT switching. In order to realize the deterministic SOT switching of the perpendicular magnetization, several methods have been developed to break the mirror symmetry, such as an external or internal in-plane magnetic field [9-11], tilted magnetic anisotropy [12], and lateral structural asymmetry or spin current asymmetry [13, 14].

Antisymmetric exchange interaction, i.e., Dzyaloshinskii–Moriya interaction (DMI) [15, 16], presents in systems without inversion symmetry, stabilizes the non-collinear spin canting between neighboring magnetic moments. DMI induces the chiral spin textures such as magnetic skyrmions [17] and the chiral magnetic coupling [18-20]. The Hamiltonian of DMI can be written as $\hat{H}_{DMI} = \sum_{i,j} -D_{ij} \cdot (m_i \times m_j)$, where $D_{ij}$ is the DMI tensor between $m_i$ and $m_j$. As a consequence of the minimized DMI energy, DMI breaks the chiral symmetry of non-collinear spin textures and determines the spin canting direction of clockwise ($D_{ij} > 0$) or anticlockwise ($D_{ij} < 0$). In the Fert-Levy model [21-23], the interfacial DMI satisfies $\boldsymbol{D}_{ij} = D_{ij}\left(\boldsymbol{r}_i \times \boldsymbol{r}_j\right)$, where $D_{ij}$ is the interfacial DMI coefficient, as shown in Fig. 1(a).

Here, we report the chiral symmetry breaking by DMI for deterministic SOT switching of the perpendicular magnetization. By introducing a gradient of saturation magnetization ($M_s$) or magnetic anisotropy field ($H_k$), non-collinear spin textures are formed as a result of the gradient of effective SOT strength, where DMI breaks the chiral symmetry of the spin textures, resulting in the deterministic SOT switching. We experimentally demonstrate this concept in three representative cases: in-plane gradient of $M_s$ in the Ta/Gd$_x$(FeCo)$_{1-x}$ system; in-plane gradient of $H_k$ in the Ta/CoFeB(wedged)/MgO system; out-of-plane gradient of $M_s$ in the Ta/[Gd$_x$(FeCo)$_{1-x}$]$_6$ and Ta/CoFeB/CoFe/MgO systems. The last case of an out-of-plane $M_s$ gradient can be applied in the wafer-scale manufacture.

Figure 1(a) shows the current-induced SOT in the heavy metal/ferromagnet system in a 20 μm×130 μm Hall bar device. A 1-ms writing current pulse $J_e$ is applied to provide the SOT firstly, followed by a 1-ms reading current pulse (1 mA) to detect the magnetization by the anomalous Hall resistance at 1-s later, where the effective field of SOT can be written as $\boldsymbol{H}_{SOT} = H_{SOT}(\boldsymbol{m} \times \boldsymbol{\sigma})$. By considering the SOT and DMI contributions, the total effective field can be written as $\boldsymbol{H}_{eff} = \boldsymbol{H}_{ex} + \boldsymbol{H}_{DMI} + \boldsymbol{H}_{SOT} + \boldsymbol{H}_k$, where $\boldsymbol{H}_{ex} = H_{ex}\Delta\boldsymbol{m}$, $\boldsymbol{H}_{DMI} = H_{DMI}\left[\left(\nabla\cdot\boldsymbol{m}\right)\hat{z} - \nabla m_z\right]$, and $\boldsymbol{H}_k = H_k\hat{z}$ represent the effective fields of exchange interaction, DMI, and perpendicular magnetic anisotropy, respectively. For the collinear initial magnetization state, $\boldsymbol{H}_{DMI}$ is equal to zero. Under the current-induced SOT, if there is a spatial gradient of $H_{eff}$, i.e., $\nabla H_{eff}$, which can be realized by the $\nabla H_{SOT}$ or $\nabla H_k$, the magnetizations at different positions along the gradient direction feel a gradient of total spin torque, and thus non-collinear spin textures ($\boldsymbol{m}_i \times \boldsymbol{m}_j$) can be formed, resulting in an DMI energy $E_{DMI}$ in the system. Figure 1(b) shows the 4 possible states for different current and magnetization configurations. For a positive DMI coefficient ($D_{ij} > 0$), such as in the Ta/Ferromagnet system [24], the clockwise chirality of spin textures is preferred due to the minimized $E_{DMI}$. The chirality of the SOT-induced spin textures is illustrated in the Fig. 1(b), where DMI breaks the chiral symmetry of these 4 possible states: for $D_{ij} > 0$ (or $D_{ij} < 0$), selects the clockwise (or anti-clockwise) chirality of ($+J_e$, $+M_z$) and ($-J_e$, $-M_z$) states [or ($+J_e$, $-M_z$) and ($-J_e$, $+M_z$) states]. Once the current is removed, the spin texture will return to its stable magnetic anisotropy direction ($\pm z$), as a result, deterministic SOT switching is achieved.

In the heavy metal/ferrimagnet [Ta/Gd$_x$(FeCo)$_{1-x}$] system, $M_s$ could be controlled by the composition of the ferrimagnet Gd$_x$(FeCo)$_{1-x}$ due to the antiferromagnetically coupled Gd and FeCo spin sublattices [25]. By co-sputtering the off-axis Gd and FeCo targets in the opposite directions with the same speed, a composition gradient is formed ($x$ from 0.14 to 0.22), where the thickness of Gd$_x$(FeCo)$_{1-x}$ film keeps uniform (Supplemental Material). As a result of this composition gradient, a saturation magnetization gradient $\nabla_y M_s$ (49 emu cm$^{-4}$) is generated along the $y$ direction, at the same time the magnetic anisotropy field $H_k$ has no obvious variation, as shown in Fig. 2(a). According to $H_{SOT} = \frac{\hbar \theta_{SH} J_e}{2|e|M_s t}$, $\nabla_y M_s$ induces the a gradient of SOT effective field $\nabla_y H_{SOT}$, where $\hbar$ is the reduced Planck constant, $t$ is the thickness of the magnetic layer, and $e$ is the elementary charge. From the Brillouin scattering (BLS) spectroscopy, we can obtain a positive $D_{ij}$ of 15.2 μJ m$^{-2}$ in this system (Supplementary Material), which stabilizes the clockwise chirality. The clockwise chirality in this system is also verified by the asymmetric magnetic domain nucleation (Supplementary Material) [26].

Then, we perform the measurement of SOT switching in the Ta/Gd$_x$(FeCo)$_{1-x}$ system with a saturation magnetization gradient $\nabla_y M_s$. For $J_e$ along the $x$ axis, due to the $\tau_{SOT}$ and $\nabla_y M_s$ along the $y$ direction, the gradient SOT-induced spin texture is formed in the $y$-$z$ plane along the $y$ direction, where DMI selects the clockwise chirality and determines the magnetization switching without external magnetic field, with a switching current density $J_c = 4.0 \times 10^6$ A cm$^{-2}$, as shown in Fig. 2(b). It is known that the chirality of spin textures could also be modified by the external magnetic field, and thus we perform the SOT switching by scanning the $H_x$, as shown in Fig. 2(c). The deterministic switching vanishes at $H_x = +30$ Oe, indicating an effective field $H_{DMI} = -30$ Oe of DMI, which is consistent with the estimated $H_{DMI} = -\frac{D}{\mu_0 M_s \delta_{DW}} = -26.4$ Oe from the BLS measurement, where $\delta_{DW} = \pi\left(\frac{A}{K}\right)^{\frac{1}{2}}$ is the domain wall width, and $A$ (4 × 10$^{-12}$ J m$^{-1}$) and $K$ (2 × 10$^4$ J m$^{-3}$) represent the exchange stiffness constant and the PMA energy, respectively [27]. The opposite polarity of SOT switching at ±$H_x$ above 60 Oe originates from the dominating contribution from the magnetic field [Fig. 2(c)].

The current-induced hysteresis loop shift method is employed to extract the out-of-plane effective field $H_z^{\text{eff}}$ of the field-free SOT for the Ta/Gd$_x$(FeCo)$_{1-x}$ system, where the center fields ($H_{zc}^+$ and $H_{zc}^-$) of $R_{xy}$-$H_z$ curves under $\pm 20$ mA are shifted to the opposite directions, as shown in Fig. 2(d). Figure 2(e) shows the $H_z^{\text{eff}} = -(H_{zc}^+ - H_{zc}^-)/2$ as a function of $J_e$, and the linear dependence shows the typical SOT characteristic.

In fact, the chiral symmetry breaking determined SOT switching is independent with the ferrimagnetic properties. Next, we demonstrate the same mechanism for a ferromagnetic CoFeB layer in the Ta/CoFeB/MgO system, where the PMA comes from the CoFeB/MgO interface and thus can be controlled by the thickness of the CoFeB layer [28]. We design a wedged CoFeB layer with the thickness ranging from 1.24 to 0.8 nm, where the CoFeB thickness has a slight gradient along the $y$ direction. The $H_k$ in devices at different positions ($y$) is measured by the saturation field of $R_{xy}$-$H_x$ curve, as shown in Fig. 3(a). $H_k$ shows the maximum at $y = 12$ mm ($t_{\text{CoFeB}} = 1.1$nm) due to the optimal Fe–O and Co–O bonds at the interface [29], where $\nabla_y H_k > 0$ in the range of $y < 12$ mm ($t_{\text{CoFeB}} > 1.1$nm), and $\nabla_y H_k < 0$ in the range of $y > 12$ mm ($t_{\text{CoFeB}} < 1.1$nm). In this case, we neglect the slight gradient of the total magnetization ($M_s t_{\text{CoFeB}}$).

Then, we perform the SOT switching in the Ta/CoFeB(wedged)/MgO system (DMI coefficient [30]: 54.0 μJ m$^{-2}$). There is a competition between $H_{\text{SOT}}$ and $H_k$: $H_{\text{SOT}}$ tilts the magnetic moment while $H_k$ brings it back to the anisotropy direction, as shown in Fig. 3(a), therefore, in the $\nabla_y H_k$ case, a smaller $H_k$ feels a larger total spin torque, i.e., a larger effective SOT strength, leading to the non-collinear spin textures. Similar to the $\nabla_y M_s$ case in Ta/Gd$_x$(FeCo)$_{1-x}$, the deterministic magnetization switching happens due to the chiral symmetry breaking of the gradient effective SOT-induced spin textures by DMI. It is worth to notice that in the devices with opposite $\nabla_y H_k$, i.e., $\nabla_y H_k > 0$ and $\nabla_y H_k < 0$, the gradient of the total effective field $\nabla_y H_{eff}$ [$\nabla_y(H_{SOT} + H_k)$, where $H_{SOT}$ is constant in this case] changes the direction, and thus reverses the SOT-induced spin canting direction between neighboring magnetic moments from $m_j \rightarrow m_i$ to $m_i \rightarrow m_j$, i.e., reverses the chirality of SOT-induced spin textures ($m_i \times m_j$). The DMI selects the same chirality (clockwise) for $\nabla_y H_k > 0$ and $\nabla_y H_k < 0$ cases, resulting in an opposite polarity of the field-free SOT switching, as shown in Fig. 3(b) and 3(c). The polarity change of the field-free SOT switching also indicates that the slightly structural mirror symmetry breaking of the wedged CoFeB layer is not the reason for deterministic switching, because it is along the same direction in all positions.

The current-induced hysteresis loop shift method is employed to analyze the relation between SOT-induced $H_z^{eff}$ and $\nabla_y H_k$, as shown in Fig. 3(d) and 3(e), where the $H_{zc}^+$ of $R_{xy}$-$H_z$ curves under the same current (+20 mA) are shifted to the opposite directions: $H_{zc}^+$ = +18 Oe for $\nabla_y H_k > 0$ and $H_{zc}^+$ = -37 Oe for $\nabla_y H_k < 0$, respectively, which is consistent with the reversed polarity of the field-free SOT switching in these two devices. We plot the $\chi_{SOT} = H_z^{eff}/J_e$ and $\nabla_y H_k$ as a function of position $y$, as shown in Fig. 3(e). The boundary between $\nabla_y H_k > 0$ region and $\nabla_y H_k < 0$ region is around $y$ = 11 mm ($t_{CoFeB}$ = 1.08 nm), where $\chi_{SOT}$ changes the sign just across this boundary, indicating the dominating contribution of $\nabla_y H_k$ on the field-free SOT. The similar results have also been observed in Ta/CoFeB/TaO$_x$(wedged) system in our previous work [13], while the macroscopic mechanism is not clear at that time, which can be well explained by our current model of chiral symmetry breaking.

The above methods of the lateral gradient lead to the variation of device properties across the wafer. In order to address this issue, we use the out-of-plane $M_s$ gradient $\nabla_z M_s$ in the multilayer stack to achieve uniform magnetic properties laterally in the wafer scale, i.e., 6 layers (totally 4 nm) of $Gd_x(FeCo)_{1-x}$, with increasing $M_s$ from 100 emu cm$^{-3}$ to 200 emu cm$^{-3}$ ($x$ from 0.21 to 0.16). In this case, the gradient SOT strength-induced spin textures are along the thickness direction ($z$), therefore, the interfacial DMI contribution in the heavy metal/ferrimagnet interface should vanish due to $E_{DMI} = \sum_{i,j} -D_{ij}(r_i \times r_j) \cdot (m_i \times m_j) = 0$. However, in the $Gd_x(FeCo)_{1-x}$ multilayer, the element distribution gradient along the $z$ direction could also give rise to an bulk (interlayer) DMI [31], which breaks the chiral symmetry and determines the SOT switching without external field, as shown in Fig. 4(a). Even for the relatively small bulk DMI ($D = 0.44$ μJ m$^{-2}$), due to $\nabla_z M_s$ is dramatically enhanced along the thickness direction ($2 \times 10^8$ emu cm$^{-4}$, $4 \times 10^6$ times than the lateral case), it can still determine the field-free SOT switching. The bulk (interlayer) DMI in multilayers has also been proved by the chiral interlayer exchange coupling phenomenon [19, 20], and the spin textures along the thickness direction have also been demonstrated [32, 33], which support our argument.

Furthermore, for the CoFeB/CoFe bilayer with a slight difference of $M_s$ (955 and 1150 eum cm$^{-3}$ for CoFeB and CoFe, respectively) in the Ta/CoFeB/CoFe/MgO structure, the chiral symmetry breaking determined SOT switching remains robust, as shown in Fig. 4(b). This structure can be directly integrated with the wafer-scale magnetic tunnel junction [34, 35] in today's magnetic memory technology.

In summary, we demonstrate that non-collinear spin textures could be formed in PMA materials under the current-induced SOT, by introducing a gradient of magnetic properties such as saturation magnetization and magnetic anisotropy field. The chiral symmetry of the SOT-induced spin textures is broken by the DMI (clockwise for $D_{ij} > 0$), resulting in the deterministic magnetization switching. The chiral symmetry breaking by DMI is topologically protected, therefore, it is robust against defects and boundaries in nanodevices.

**Acknowledgements:**

This work is supported by the NSF Award Nos. 1611570 and 1619027, the Nanosystems Engineering Research Center for Translational Applications of Nanoscale Multiferroic Systems (TANMS), the U.S. Army Research Office MURI program under Grants No. W911NF-16-1-0472 and No. W911NF-15-1-10561, and the Spins and Heat in Nanoscale Electronic Systems (SHINES) Center funded by the US Department of Energy (DOE), under Award No. DE-SC0012670. D.L and X. L. are funded by an NSF MRSEC under Cooperative Agreement No. DMR-1720595. We are also grateful to the support from the Function Accelerated nanoMaterial Engineering (FAME) Center, and a Semiconductor Research Corporation (SRC) program sponsored by Microelectronics Advanced Research Corporation (MARCO) and Defense Advanced Research Projects Agency (DARPA).

# Figures:

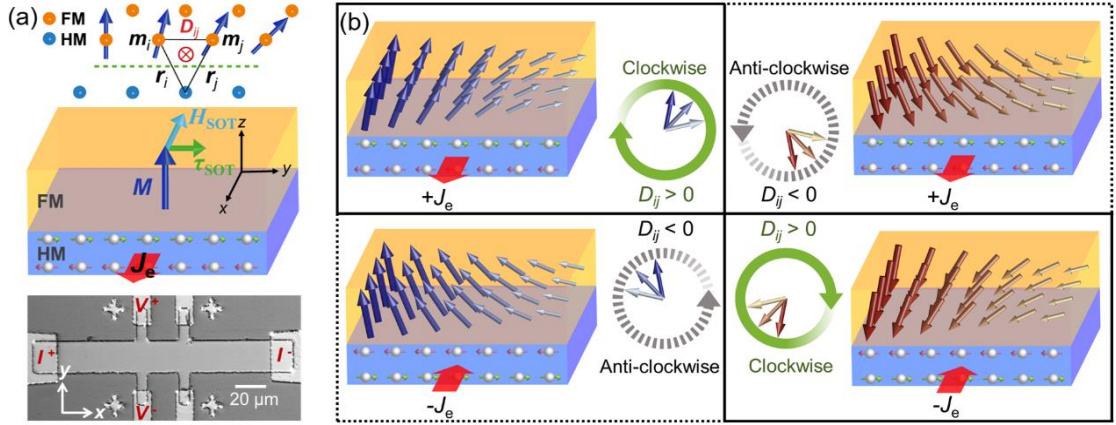

FIG. 1. The concept of deterministic SOT switching by chiral symmetry breaking. (a) The spin current in the heavy metal (HM) generated by spin Hall effect exerts a torque $\tau_{SOT}$ on the adjacent magnet (FM) with perpendicular magnetic anisotropy (PMA), which can be described by an effective field $H_{SOT}$, where $J_e$ is the (electric) charge current. The schematic of the interfacial Dzyaloshinskii–Moriya interaction (DMI) ($D_{ij} > 0$), which stabilizes the clockwise chirality. Bottom figure shows the experiment set up in the Hall bar device. (b) Under the current-induced spin-orbit torque (SOT), if there is a gradient of the total effective field $H_{eff}$, the non-collinear spin texture ($m_i \times m_j$) is formed, resulting in an DMI energy in the system. There are 4 possible states for different current and magnetization configurations, where the chiral symmetry is broken by the DMI ($D_{ij} > 0$ for clockwise or $D_{ij} < 0$ for anti-clockwise), leading to the deterministic magnetization switching.

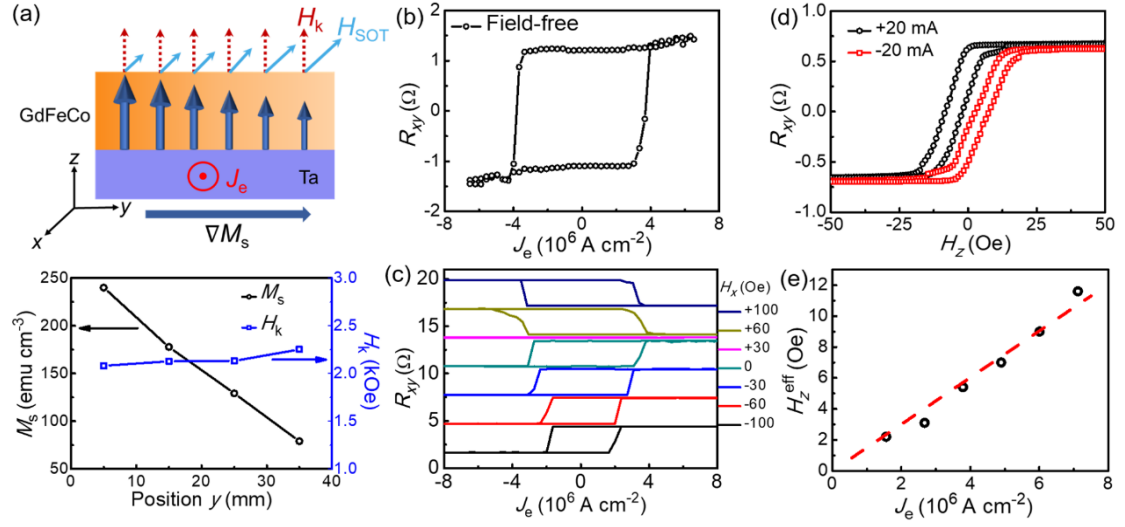

FIG. 2. Chiral symmetry breaking determined SOT switching with an in-plane gradient of $M_s$ ($\nabla_y M_s$). (a) The composition gradient along the $y$ direction in Ta/Gd$_x$(FeCo)$_{1-x}$ leads to a saturation magnetization gradient $\nabla_y M_s$ (49 emu cm$^{-4}$), where the magnetic anisotropy field $H_k$ is almost constant. (b) Current-induced SOT switching without the external magnetic field. (c) SOT switching with a series of in-plane magnetic field $H_x$. (d) $R_{xy}$-$H_z$ curves with ±20 mA current show the field-free SOT-induced hysteresis loop shift. (e) Out-of-plane effective field $H_z^{\text{eff}}$ of field-free SOT as a function of $J_e$.

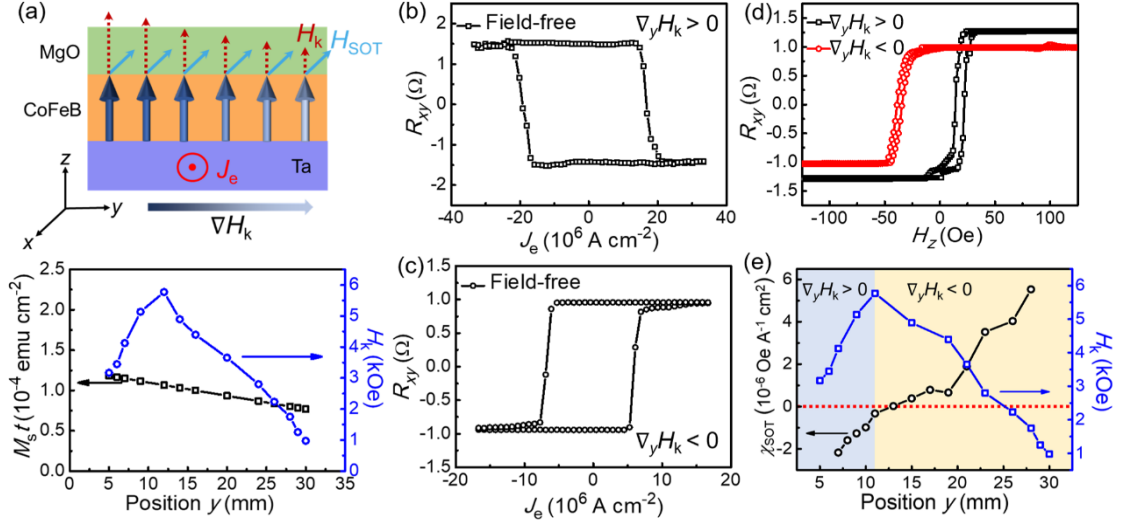

FIG. 3. Chiral symmetry breaking determined SOT switching with an in-plane gradient of $H_k$ ($\nabla_y H_k$). (a) In the Ta/CoFeB(wedged)/MgO system, the magnetic anisotropy field $H_k$ from the interfacial PMA could be modulated by varying the thickness of CoFeB, resulting in a $\nabla_y H_k$ along the $y$ direction. (b) and (c) show the field-free SOT switching in $\nabla_y H_k > 0$ and $\nabla_y H_k < 0$ cases, respectively, where the switching polarity is opposite in these two cases. (d) Field-free SOT-induced hysteresis loops are shifted to the opposite field directions in $\nabla_y H_k > 0$ and $\nabla_y H_k < 0$ cases under the same current (20 mA). (e) Field-free SOT-induced out-of-plane effective field $\chi_{SOT}$ as a function of device positions, where $\chi_{SOT}$ changes the sign from $\nabla_y H_k > 0$ region to $\nabla_y H_k < 0$ region.

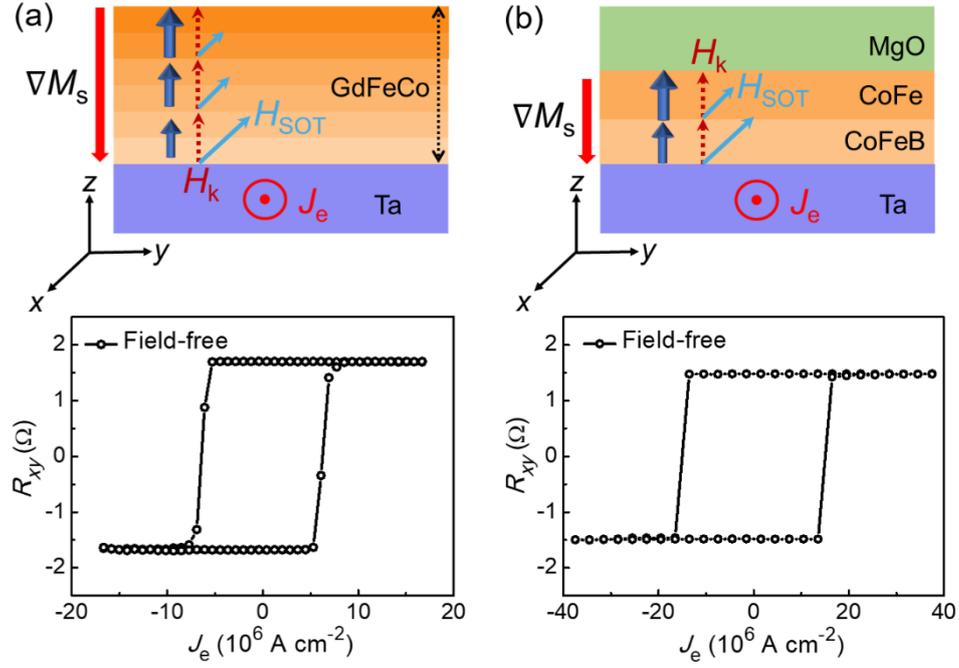

FIG. 4. Chiral symmetry breaking determined SOT switching with an out-of-plane gradient of $M_s$ ($\nabla_z M_s$). The chiral symmetry breaking determined SOT switching is robust in the ferrimagnet-based Ta/[Gd$_x$(FeCo)$_{1-x}$]$_6$ system with 6 magnetic layers with increasing $M_s$ (a) and in the ferromagnet-based Ta/CoFeB/CoFe/MgO system with a magnetic bilayer (b), where the $\nabla_z M_s$ exists along the thickness ($z$) direction.

**References:**

[1]  P. W. Anderson, Science **177** (4047), 393-396 (1972).
[2]  L. Liu, C.-F. Pai, Y. Li, H. W. Tseng, D. C. Ralph and R. A. Buhrman, Science **336** (6081), 555-558 (2012).
[3]  L. Liu, O. J. Lee, T. J. Gudmundsen, D. C. Ralph and R. A. Buhrman, Physical Review Letters **109** (9), 096602 (2012).
[4]  I. M. Miron, K. Garello, G. Gaudin, P.-J. Zermatten, M. V. Costache, S. Auffret, S. Bandiera, B. Rodmacq, A. Schuhl and P. Gambardella, Nature **476**, 189 (2011).
[5]  J. Sinova, S. O. Valenzuela, J. Wunderlich, C. H. Back and T. Jungwirth, Reviews of Modern Physics **87** (4), 1213-1260 (2015).
[6]  A. Manchon, H. C. Koo, J. Nitta, S. M. Frolov and R. A. Duine, Nature Materials **14**, 871 (2015).
[7]  S. Ikeda, K. Miura, H. Yamamoto, K. Mizunuma, H. D. Gan, M. Endo, S. Kanai, J. Hayakawa, F. Matsukura and H. Ohno, Nature Materials **9**, 721 (2010).
[8]  B. Dieny and M. Chshiev, Reviews of Modern Physics **89** (2), 025008 (2017).
[9]  Y.-C. Lau, D. Betto, K. Rode, J. M. D. Coey and P. Stamenov, Nature Nanotechnology **11**, 758 (2016).
[10] Y.-W. Oh, S.-h. Chris Baek, Y. M. Kim, H. Y. Lee, K.-D. Lee, C.-G. Yang, E.-S. Park, K.-S. Lee, K.-W. Kim, G. Go, J.-R. Jeong, B.-C. Min, H.-W. Lee, K.-J. Lee and B.-G. Park, Nature Nanotechnology **11**, 878 (2016).
[11] S. Fukami, C. Zhang, S. DuttaGupta, A. Kurenkov and H. Ohno, Nature Materials **15**, 535 (2016).
[12] L. You, O. Lee, D. Bhowmik, D. Labanowski, J. Hong, J. Bokor and S. Salahuddin, Proceedings of the National Academy of Sciences **112** (33), 10310-10315 (2015).
[13] G. Yu, P. Upadhyaya, Y. Fan, J. G. Alzate, W. Jiang, K. L. Wong, S. Takei, S. A. Bender, L.-T. Chang, Y. Jiang, M. Lang, J. Tang, Y. Wang, Y. Tserkovnyak, P. K. Amiri and K. L. Wang, Nature Nanotechnology **9**, 548 (2014).
[14] K. Cai, M. Yang, H. Ju, S. Wang, Y. Ji, B. Li, K. W. Edmonds, Y. Sheng, B. Zhang, N. Zhang, S. Liu, H. Zheng and K. Wang, Nature Materials **16**, 712 (2017).
[15] I. Dzyaloshinsky, Journal of Physics and Chemistry of Solids **4** (4), 241-255 (1958).
[16] T. Moriya, Physical Review **120** (1), 91-98 (1960).
[17] A. Fert, N. Reyren and V. Cros, Nature Reviews Materials **2**, 17031 (2017).
[18] Z. Luo, T. P. Dao, A. Hrabec, J. Vijayakumar, A. Kleibert, M. Baumgartner, E. Kirk, J. Cui, T. Savchenko, G. Krishnaswamy, L. J. Heyderman and P. Gambardella, Science **363** (6434), 1435-1439 (2019).
[19] D.-S. Han, K. Lee, J.-P. Hanke, Y. Mokrousov, K.-W. Kim, W. Yoo, Y. L. W. van Hees, T.-W. Kim, R. Lavrijsen, C.-Y. You, H. J. M. Swagten, M.-H. Jung and M. Kläui, Nature Materials **18** (7), 703-708 (2019).
[20] A. Fernández-Pacheco, E. Vedmedenko, F. Ummelen, R. Mansell, D. Petit and R. P. Cowburn, Nature Materials **18** (7), 679-684 (2019).
[21] A. Fert and P. M. Levy, Physical Review Letters **44** (23), 1538-1541 (1980).
[22] P. M. Levy and A. Fert, Physical Review B **23** (9), 4667-4690 (1981).
[23] H. Yang, A. Thiaville, S. Rohart, A. Fert and M. Chshiev, Physical Review Letters **115** (26), 267210 (2015).
[24] X. Ma, G. Yu, C. Tang, X. Li, C. He, J. Shi, K. L. Wang and X. Li, Physical Review Letters **120**

(15), 157204 (2018).

[25] I. Radu, K. Vahaplar, C. Stamm, T. Kachel, N. Pontius, H. A. Dürr, T. A. Ostler, J. Barker, R. F. L. Evans, R. W. Chantrell, A. Tsukamoto, A. Itoh, A. Kirilyuk, T. Rasing and A. V. Kimel, Nature **472**, 205 (2011).

[26] S. Pizzini, J. Vogel, S. Rohart, L. D. Buda-Prejbeanu, E. Jué, O. Boulle, I. M. Miron, C. K. Safeer, S. Auffret, G. Gaudin and A. Thiaville, Physical Review Letters **113** (4), 047203 (2014).

[27] L. Caretta, M. Mann, F. Büttner, K. Ueda, B. Pfau, C. M. Günther, P. Hessing, A. Churikova, C. Klose, M. Schneider, D. Engel, C. Marcus, D. Bono, K. Bagschik, S. Eisebitt and G. S. D. Beach, Nature Nanotechnology **13** (12), 1154-1160 (2018).

[28] M. Endo, S. Kanai, S. Ikeda, F. Matsukura and H. Ohno, Applied Physics Letters **96** (21), 212503 (2010).

[29] H. X. Yang, M. Chshiev, B. Dieny, J. H. Lee, A. Manchon and K. H. Shin, Physical Review B **84** (5), 054401 (2011).

[30] X. Ma, G. Yu, X. Li, T. Wang, D. Wu, K. S. Olsson, Z. Chu, K. An, J. Q. Xiao, K. L. Wang and X. Li, Physical Review B **94** (18), 180408 (2016).

[31] D.-H. Kim, M. Haruta, H.-W. Ko, G. Go, H.-J. Park, T. Nishimura, D.-Y. Kim, T. Okuno, Y. Hirata, Y. Futakawa, H. Yoshikawa, W. Ham, S. Kim, H. Kurata, A. Tsukamoto, Y. Shiota, T. Moriyama, S.-B. Choe, K.-J. Lee and T. Ono, Nature Materials **18** (7), 685-690 (2019).

[32] W. Legrand, J.-Y. Chauleau, D. Maccariello, N. Reyren, S. Collin, K. Bouzehouane, N. Jaouen, V. Cros and A. Fert, Science Advances **4** (7), eaat0415 (2018).

[33] W. Li, I. Bykova, S. Zhang, G. Yu, R. Tomasello, M. Carpentieri, Y. Liu, Y. Guang, J. Gräfe, M. Weigand, D. M. Burn, G. van der Laan, T. Hesjedal, Z. Yan, J. Feng, C. Wan, J. Wei, X. Wang, X. Zhang, H. Xu, C. Guo, H. Wei, G. Finocchio, X. Han and G. Schütz, Advanced Materials **31** (14), 1807683 (2019).

[34] S. S. P. Parkin, C. Kaiser, A. Panchula, P. M. Rice, B. Hughes, M. Samant and S.-H. Yang, Nature Materials **3** (12), 862-867 (2004).

[35] S. Yuasa, T. Nagahama, A. Fukushima, Y. Suzuki and K. Ando, Nature Materials **3** (12), 868-871 (2004).